\documentclass[
superscriptaddress,twocolumn,
 amsmath,amssymb,
 aps,
prd,
]{revtex4-1}

\usepackage{tensor}

\usepackage{xcolor}

\usepackage{hyperref}

\usepackage{appendix}

\usepackage{graphicx}
\usepackage{dcolumn}
\usepackage{bm} 

\def\bea {\begin{eqnarray}}
\def\eea {\end{eqnarray}}
\def\nn {\nonumber}
\def\p {\partial}

\begin{document}

\title{General covariance and dynamics with a Gauss law}
  
\author{Viqar Husain} \email{vhusain@unb.ca}
\affiliation{Department of Mathematics and Statistics, University of New Brunswick, Fredericton, NB Canada E3B 5A3 }
\affiliation{Perimeter Institute for Theoretical Physics,
 31 Caroline Street North, Waterloo ON N2L 2Y5, Canada }
 
 \author{Hassan Mehmood}\email{hassan.mehmood@unb.ca}
 \affiliation{Department of Mathematics and Statistics, University of New Brunswick, Fredericton, NB Canada E3B 5A3 }
 
\date{\today} 

\begin{abstract}
We present a 4-dimensional generally covariant gauge theory with local degrees of freedom which leads to the Gauss constraint but lacks {\it both} the Hamiltonian and spatial diffeomorphism constraints. The canonical theory  therefore resembles Yang-Mills theory without the Hamiltonian. We describe its observables, quantization, and some generalizations. 

\end{abstract}

\maketitle


\section{\label{sec1} Introduction}

The difficulty of constructing a fully satisfactory theory of quantum gravity gives rise to the need for simpler models, which consist in writing down generally covariant theories that are different from general relativity but simpler to analyze from a quantum-theoretical perspective. These include BF theory \cite{Horowitz:1989ng,Husain:1990sc}, 2+1 gravity \cite{Witten:1988hc,Ashtekar:1989qd,Carlip:1989nz,Moncrief:1990mk},  $U(1)^3$ theory \cite{Smolin:1992wj,Bakhoda:2020ril}, the Husain-Kuchar (HK) model \cite{1990PhRvD..42.4070H,BarberoG:2000dsc}, and many others with symmetry reductions \cite{BarberoG:2010oga} and matter. 

The Hamiltonian version of the HK theory lacks the Hamiltonian constraint of general relativity, which encodes the dynamics of general relativity and is the most difficult object to quantize in any canonical approach to quantum gravity. As such, the model furnished the hope for clearly separating the question of dynamics from that of kinematics, and thus determining which quantum-gravitational effects are solely kinematical in nature. To this end, subsequent developments revealed subtle and surprising results, such as the fact that black-hole entropy calculations in loop quantum gravity (LQG) can be performed with generally covariant theories that lack black-hole solutions entirely \cite{PhysRevD.59.084019}. 

In this paper, we describe an SU(2) model that, on the face of it, is a slight variation of the HK model, but turns out to have significantly different features: the canonical theory lacks not only the Hamiltonian, but also the diffeomorphism constraint; the remaining Gauss constraint arises with a source term. This leads to new observables in the classical and quantum theories that resemble ``quarks on a string." Furthermore, the presence of only the Gauss constraint gives considerable freedom in constructing  classical solutions, which include metrics corresponding to almost any three-geometry.  

The surprising feature that a generally covariant theory realizes dynamics only as internal gauge provides a large class of exactly solvable classical and quantum models with local degrees of freedom; the canonical theory of the model we present may be viewed as a Yang-Mills theory without its Hamiltonian, with nine local degrees of freedom. 

The plan of the paper is as follows. In Section II, we introduce the model and derive its canonical theory, and in section III, we explain why the model is diffeomorphism-invariant despite having a Gauss law as the only first-class constraint and hence gauge generator of the theory. Sections IV and V are devoted to describing certain classical observables and solutions, respectively, and Section VI discusses the quantum theory. Finally, Section VII describes possible extensions of the model, such as addition of a Chern-Simons boundary term.

\section{\label{sec2} Action and canonical theory}
Let $M$ be a four-dimensional spacetime, and let $\phi^i(x)$ and $A^i_\alpha(x)$ be scalars and connection one-forms in the Lie algebra $\mathfrak{su}(2)$ of $SU(2)$ (or any other semisimple \footnote{The condition of being semisimple is necessary, for by Cartan's criterion, the Lie algebra of a group is semisimple if and only if it possesses a non-degenerate Cartan-Killing metric, which is required to form a totally antisymmetric object (e.g. the Levi-Civita tensor in $\mathfrak{su}(2)$ from the structure constants of the Lie algebra.)} Lie group of dimension 3, e.g. $SL(2,\mathbb{R})$, $SO(2,1)$, etc.), and let $\epsilon_{ijk}$ denote the Levi-Civita symbol in $\mathfrak{su}(2)$ (or equivalently, the structure constants of the Lie algebra). The action of the model we propose is 
\begin{align}
    S[\phi, A] &= \frac{1}{2}\int_M d^4x\ \tilde{\epsilon}^{\alpha\beta\gamma\delta}\epsilon_{ijk}D_\alpha\phi^i D_\beta\phi^j F^k_{\gamma\delta}  \nn\\
    &= \frac{1}{2}\int_M d^4x\,\text{Tr}\,(D\phi \wedge D\phi \wedge F),
    \label{act2}
\end{align}
where $D$ and $F$ denote, respectively, the covariant derivative and the curvature 2-form corresponding to the connection $A$, i.e.
\begin{eqnarray}
    D_\alpha\gamma^i &= \partial_\alpha\gamma^i + \tensor{\epsilon}{^i_{jk}}A^j_{\alpha}\gamma^k \\
    F^i_{\alpha\beta} &= \partial_{[\alpha}A^i_{\beta]} + \tensor{\epsilon}{^i_{jk}}A^j_{\alpha}A^k_{\beta},
\end{eqnarray}
and $\tilde{\epsilon}^{\alpha\beta\gamma\delta}$ is the four-dimensional densitized Levi-Civita symbol. 

This action  is manifestly invariant both under spacetime diffeomorphisms and $SU(2)$ transformations. It resembles the HK action 
\bea
S_{\rm HK}[e,A] = \int d^4x\ \tilde{\epsilon}^{\alpha\beta\gamma\delta}\epsilon_{ijk}e_\alpha^i e_\beta^j F_{\gamma\delta}^k
\eea
if one replaces $D_\alpha\phi^i$ with $SU(2)$ triads $e_\alpha^i$. However, as we see below, the canonical theory of (\ref{act2}) is drastically different.

Variation of the action with respect to $A$ and $\phi$ yields the equations of motion:
\begin{align}
    \delta\phi&: \qquad\qquad D \wedge (D\phi \wedge F) = 0, \label{deltaphi}\\
    \delta A&: \qquad D \wedge (D\phi \wedge D\phi) + \phi \times (D\phi \wedge F) = 0 \label{deltaA},
\end{align}
where `$\times$' denotes an internal cross product using the structure constants, i.e. $(u\times v)^i = \tensor{\epsilon}{^i_{jk}}u^jv^k$. Using the Bianchi identity $D_{[\alpha} F_{\beta\gamma]} = 0$ and $D_{[\alpha}D_{\beta]}\lambda = F_{\alpha\beta}\times\lambda$ for any section $\lambda$ of the $SU(2)$ bundle associated with the spacetime, the preceding equations can be recast as
\begin{eqnarray}
    \tilde{\epsilon}^{\alpha\beta\gamma\delta}(F_{\alpha\beta}\times \phi)&\times& F_{\gamma\delta} = 0, \nonumber \\
    \tilde{\epsilon}^{\alpha\beta\gamma\delta}[(F_{\beta\gamma}\times\phi)\times D_{\delta}\phi &+&\phi\times(D_\beta\phi\times F_{\gamma\delta})] = 0.\nonumber
\end{eqnarray}
Evidently, the first equation is true if and only if $F$ and $\phi$ are internally parallel, i.e. 
\bea
F_{\alpha\beta} \times \phi = 0.
\label{fcrossphi}
\eea
Substituting this equation and its covariant derivative into the second equation trivially satisfies the latter. Thus, any connection is a solution to the theory, provided one chooses a section $\phi$ parallel to the curvature $F$ in $\mathfrak{su}(2)$! One might say that the equations of motion concern only the ``internal space" of the theory; from this perspective, it is not surprising that the canonical theory, as we shall see below, only has an internal Gauss constraint. 

To perform canonical decomposition of the action (\ref{act2}), we assume $M$ has the topology $\mathbb{R}\times\Sigma$, where $\Sigma$ is a compact three-dimensional, differentiable manifold. Then, choosing the  coordinates $x^\alpha = (x^0, x^a)$, with $x^0 \in \mathbb{R}$ and $x^a\in\Sigma$, the $3+1$ form of the action is
\bea
S &=& \int d^4x\ \epsilon_{ijk}\ \tilde{\epsilon}^{0abc} \left\{\left(\dot{\phi}^i + A_0^i\epsilon^i_{\ lm} A_a^l\phi^m\right)D_a\phi^j F_{bc}^k  \right.\nn\\
&& + \left.   D_a\phi^i D_b \phi^j\left( \dot{A}_c^k + \epsilon^k_{\ lm}A_0^lA_c^m \right) \right\},
\eea
where the overdot indicates $\partial_0$. This identifies the momenta conjugate to $\phi^i$ and $A_a^i$:
\begin{eqnarray}
    \tilde{E}^a_i &:=& \tilde{\epsilon}^{abc}\epsilon_{ijk}(D_b\phi^j) (D_c\phi^k), \label{E} \label{eq9}\\
    \tilde{p}_i &:=& \tilde{\epsilon}^{abc}\epsilon_{ijk}(D_{a}\phi^j) F^k_{bc} \label{p}\label{eq10},
\end{eqnarray}
where $\tilde{\epsilon}^{abc}:= \tilde{\epsilon}^{0abc}$. The canonical action is then  
\begin{equation}
    S = \int d^4x \left[\tilde{E}^c_k\dot{A}^k_c + \tilde{p}_i\dot{\phi}^i + A^k_0(D_c\tilde{E}^c_k + \tensor{\epsilon}{_{kl}^i}\phi^l\tilde{p}_i) \right] \label{eq11}
 \end{equation}
 There are thus two sets of configuration variables, the $\mathfrak{su}(2)$-valued connection $A$ and scalar $\phi$. Variation with respect to $A_0^k$ gives the constraint
\begin{equation}
    \tilde{G}_k = -(D_c\tilde{E}^c_k + \tensor{\epsilon}{_{kl}^i}\phi^l\tilde{p}_i) \approx 0 \label{gauss}.
\end{equation}
This is a Gauss law with a source term; it is readily verified that the constraint algebra is first class: 
\begin{widetext}
    \bea
    \{G(\lambda), G(\mu)\} &=& \left\{\int d^3x\ \lambda^mD_a\tilde{E}^a_m(x),  \int d^3y\ \gamma^nD_b\tilde{E}^b_n(y)\right\} 
    + \left\{\int d^3x\ \lambda^m\tensor{\epsilon}{_{ml}^s}\phi^l\tilde{p}_s(x), \int d^3y\ \gamma^n\tensor{\epsilon}{_{ij}^k}\phi^j\tilde{p}_k(y)\right\}\nn\\
    &=& G([\lambda,\mu]) \label{eq13}.
    \eea
\end{widetext}
The Hamiltonian is therefore a linear combination of first class constraints 
\begin{eqnarray}
    H = \int d^3x \,\lambda_k G^k(x),
\end{eqnarray}
and it is evident that there is neither a Hamiltonian nor a spatial diffeomorphism constraint; the reason for their absence is explored in the next section. 

At this stage it is important to check whether any  constraints have been missed. We now show that this is not the case. Let us first note that there can be no algebraic relations between $\phi^i$, $A_a^i$ and $E^{ai}$, since the former two are independent configuration variables, and the last one is made from the gauge covariant derivatives of the first (\ref{E}); similarly the momenta $\tilde{E}^{ai}$ and $\tilde{p}^i$ are algebraically independent since the latter, from (\ref{p}), is a function of the curvature $F(A)$; the possible nontrivial quadratic combinations of the momenta (with $B^{ai} = \tilde{\epsilon}^{abc} F_{ab}^i$) are  
\begin{eqnarray}
     \Tilde{p_i}\Tilde{p}^i =  B^{aj}B^b_jD_a\phi^iD_b\phi_i - B^{aj}B^b_iD_a\phi^iD_b\phi_j \\
     \Tilde{E}^{ai}\Tilde{E}^b_i = 2\Tilde{\epsilon}^{acd}\Tilde{\epsilon}^{bef}D_c\phi^jD_d\phi^kD_e\phi_jD_f\phi_k \\
     (\Tilde{E}^a\times \Tilde{E}^b)_i = \tensor{\epsilon}{_i^{jk}}\Tilde{E}^a_j\Tilde{E}^b_k =2\Tilde{E}^a_j\Tilde{\epsilon}^{bcd}D_c\phi_iD_d\phi^j\\
     (\Tilde{p}\times\Tilde{E}^a)_i = \tensor{\epsilon}{_i^{jk}}\Tilde{p}_j\Tilde{E}^a_k = 2\Tilde{p}_j\Tilde{\epsilon}^{abc}D_b\phi_iD_c\phi^j;
 \end{eqnarray}
 inspecting the right hand sides of these expressions shows that there can be no quadratic relationships between the momenta. We thus conclude that there are no ``hidden" secondary constraints in the model. Another way to establish this via the Dirac procedure for constrained Hamiltonian systems \cite{dirac2001lectures, Henneaux:1992ig} is given in the appendix. 

The theory thus has three $\phi^i$ and nine $A_a^i$ local phase space configuration degrees of freedom subject to the $SU(2)$ Gauss law. Hence it has a net of nine unconstrained degrees of freedom per space point \cite{Henneaux:1992ig}; in comparison the HK model has three, since it has only the connection $A_a^i$ as the configuration variable, and both the Gauss and diffeomorphism constraints.
 
Like the HK model, the action (\ref{act2}) has a special vector density
\bea
\tilde{u}^\alpha = \epsilon^{\alpha\beta\gamma\delta} \epsilon_{ijk}D_\beta\phi^i D_\gamma\phi^j D_\delta\phi^k \label{ut}
\eea
which satisfies the ``parallel transport"  equation $\tilde{u}^\alpha D_\alpha\phi^i=0$; a degenerate spacetime metric 
\bea
g_{\alpha\beta}= D_\alpha\phi^i D_\beta\phi^j\delta_{ij};
\label{deg-m}
\eea
in the canonical theory there is a scalar density of weight one
\bea
\tilde{e} = \frac{1}{3!}\epsilon_{ijk}\tilde{\epsilon}^{abc}D_a\phi^iD_b\phi^jD_c\phi^k \label{ed}
\eea
which may be used to define the inverse triad 
\bea
e^{ai} = \frac{1}{\tilde{e}}\epsilon^{ijk} \tilde{\epsilon}_{abc}D_b\phi^jD_c\phi^k,  \label{inve}
\eea
 which satisfies $e^{ai}e_{aj} = \delta^i_j$ and $e^{ai}e_{bi} = \delta^a_b$. Thus like the HK model, (\ref{act2}) is a theory of $3-$geometries with an invertible spatial metric $q_{ab}=e_a^ie_b^i$.

 \section{The case of the missing constraints}
 
 The action \eqref{act2} is manifestly diffeomorphism invariant, but the canonical theory contains only a Gauss constraint. Since one expects continuous symmetries of the action to be generated by the first-class constraints, we have a puzzle: how do we understand the absence of the diffeomorphism and Hamiltonian constraints?  

By way of preliminaries, let us recall how this question is answered in other known diffeomorphism-invariant theories of connections where the first-class constraints generate only the usual gauge transformations of the gauge fields. Examples include $BF$ theory and its generalizations \cite{Horowitz:1989ng,Husain:1990sc}, Chern-Simons theory, and $2+1$ gravity \cite{Witten:1988hc}. In these theories $F=0$ is an equation of motion. As a result the Lie derivative of the field variables with respect to a vector field $\zeta$ that generates a diffeomorphism is a gauge transformation on shell:
\bea
{\cal L}_\zeta A_a &=& \zeta^c\p_c A_a + A_c\p_a\zeta^c\nn \label{eq23}\\
&=& D_a(\zeta^cA_c) + \zeta^cF_{ca}. 
\label{LieA}
\eea
The reflection of this fact in the canonical BF theory arises through the constraints
$G^k =D_aE^{ak}=0$ and $F_{ab}^k=0$ with the spatial diffeomorphism constraint $C_a$ arising as the linear combination
\bea
C_a = A_a^kG^k + E^{ak}F_{ab}^k
\eea

In the HK model, the Gauss and spatial diffeomorphism constraints are present but the Hamiltonian constraint vanishes identically. Although the latter cannot be written as any linear combination of the former constraints, the Lie-derivative argument sketched above is still available: the theory contains the vector density 
$\tilde{u}^a \equiv \epsilon^{abcd} e_b^ie_c^je_d^k\epsilon^{ijk}$, which  defines a preferred direction, and satisfies $\tilde{u}^aF_{ab}=0$ \cite{1990PhRvD..42.4070H}. Using this, one can show, in analogy with the theories discussed above, that spacetime diffeomorphisms are equivalent on shell to the transformations generated by the first-class constraints of the theory, namely the Gauss law and the spatial diffeomorphism constraint. Indeed, given a foliation of spacetime into a family of spacelike hypersurfaces, one can convert $\tilde{u}$ into a vector field $u$, and then any vector field $\zeta$ can be decomposed into a component along $u$ and a component along a leaf $\Sigma$ of the foliation. Then, up to unimportant multiplicative factors, one can write
\bea
    \mathcal{L}_{\zeta}A_a &=&  D_a(u^bA_b) + u^bF_{ba} + \mathcal{L}_{X}A_a \nonumber \\
    \mathcal{L}_{\zeta}e_a &=& (u^bA_b)\times e_a + u^bD_b e_a + \mathcal{L}_Xe_a
\eea
where $X$ is a vector field along $\Sigma$. The first and last terms in each line are an $SU(2)$ rotation and a spatial diffeomorphism, respectively, and the middle term vanishes by $\tilde{u}^aF_{ab} = 0$, which is a consequence of the equations of motion \cite{1990PhRvD..42.4070H}. Furthermore, one can show that the spatial projections of the equations of motion yield the Gauss and spatial diffeomorphism constraints. Thus although the HK model is not topological with an $F_{ab}=0$ equation of motion, a related understanding of the identically vanishing Hamiltonian constraint arises there.

We now show that there is an analogous understanding of the vanishing Hamiltonian and diffeomrophism constraints in the canonical theory arising out of the action (\ref{act2}).  That is, we can show that all diffeomorphisms are equivalent on shell to the $SU(2)$ gauge transformations generated by the Gauss law \eqref{gauss}. To see how this happens let us recall the special vector density $\tilde{u}^\alpha$ (\ref{ut}) orthogonal to $D_\alpha\phi$;  it can be converted into a vector field using \eqref{ed}. Furthermore, in the canonical theory, one has access to the three vector fields $e^{ai}$ \eqref{inve} which have the property that $e^{ai}D_a\phi^j = \delta^{ij}$. Using this, let us form the spatial vector field
\begin{equation}
    X^{ai} = \tensor{\epsilon}{^i_{jk}}\phi^je^{ak}
\end{equation}
Any vector field $v^\alpha$ can be decomposed into a component along $u^\alpha=\tilde{u}^\alpha/\tilde{e}$ and a spatial vector field, which in turn can be written as a linear combination of the $X^{ai}$. Thus, it suffices to calculate the Lie derivative of the field variables along $u^\alpha$, and along $w^a = \lambda_iX^{ai}$ for some $\lambda_i$. To this end, we first note that
\begin{eqnarray}
    \mathcal{L}_{u}\phi^i = u^\alpha \partial_\alpha \phi^i &=& u^\alpha D_\alpha\phi - [(u^\alpha A_\alpha) \times \phi]^i \nonumber \\
    &=& - [(u^\alpha A_\alpha)\times\phi]^i,
\end{eqnarray}
where the last equality follows from $u^\alpha D_\alpha \phi = 0$---this is evidently an $SU(2)$ gauge transformation with the gauge parameter $u^\alpha A_\alpha^i$. Similarly, for the Lie derivative with respect to $w^a$,  we find  
\bea
    \mathcal{L}_{w}\phi^i &=& [(\lambda - (A_a w^a))\times\phi]^i\nn\\
    &\equiv& \left(\Lambda \times \phi\right)^i
    \label{Lwphi}
\eea
which is also an $SU(2)$ gauge transformation with the parameter $\Lambda^i$. Therefore, by virtue of the special direction $u^\alpha$, any diffeomorphism of $\phi$ is equivalent to an $SU(2)$ rotation. 

This fact can be used to establish a similar result for $A_a^i$, provided the equations of motion hold. To see this, observe that if the equations of motion hold, then as shown in the previous section, $F\times\phi = 0$. Now, under a diffeomorphism generated by a vector field $v^\alpha$, $\phi$ changes by an $SU(2)$ rotation. Since $F\times\phi = 0$, $F$ must rotate by the same amount as $\phi$ does \eqref{Lwphi}. Indeed, assuming that $\phi$ rotates by $\Lambda^i$,
\begin{eqnarray*}
    \mathcal{L}_v(F_{\alpha}\times\phi) = \mathcal{L}_vF_{\alpha\beta}\times\phi + F_{\alpha\beta}\times(\Lambda\times\phi) = 0;
\end{eqnarray*}
this equation holds if and only if 
\bea
\mathcal{L}_{v}F_{\alpha\beta} = (\Lambda\times F_{\alpha\beta}),
\label{LieF}
\eea
i.e. $F$, too, rotates by $\Lambda$. This constrains the transformation of the connection $A$: we have, using \eqref{LieA}, that 
\bea
\mathcal{L}_v F_{\alpha\beta} &=& D_{[\alpha}\mathcal{L}_v A_{\beta]}\nn\\
&=& (F_{\alpha\beta}\times A_\gamma v^\gamma) - D_{[\alpha}(v^\gamma F_{\beta]\gamma});
\eea
and the last two equations give
\bea
D_{[\alpha}(v^\gamma F_{\beta]\gamma}) &=& F_{\alpha\beta}\times(A_\gamma v^\gamma + \Lambda).
\eea
This in turn implies 
\bea
v^\gamma F_{\alpha\gamma} = D_\alpha (A_\gamma v^\gamma + \Lambda);
\eea
substituting this into \eqref{LieA} gives
\begin{equation}
    \mathcal{L}_v A_\alpha^i = -D_\alpha \Lambda^i;
\end{equation}
this is the sought after result: the Lie derivative of $A$ is an $SU(2)$ gauge rotation. Therefore, provided the equations of motion hold, diffeomorphisms are equivalent to the transformations generated by the Gauss constraint. To complete the analogy with the HK model, one can  project eqn. \eqref{deltaA} into the spatial surface to obtain the Gauss constraint \eqref{gauss}; (\eqref{deltaphi} is a scalar and hence not projectable). 

From the perspective of constrained Hamiltonian systems, the preceding discussion illustrates the fact that all continuous local symmetries of the action are generated by the first-class constraints; any transformations that are not present in the transformations generated by the first-class constraints should differ from the latter only by equations-of-motion terms or by trivial symmetries \cite{Henneaux:1992ig}. This is what happens here.

\section{\label{sec3} Observables}
Let us recall that an observable is a dynamical function(al) $f(\phi, A, \tilde{E}, \tilde{p})$ that has vanishing Poisson brackets with the constraints. We can define a spatial metric by 
\begin{equation}
    q_{ab} = \kappa_{ij}D_a\phi^i D_b\phi^j \label{space-m};
\end{equation}
where $\kappa_{ij}$ is the Cartan-Killing metric for $SU(2)$. It is readily verified that this is an observable: 
\begin{equation}
    \{D_a\phi^i(x), G(\lambda)\}=\tensor{\epsilon}{^i_{mn}}\lambda^mD_a\phi^n,
\end{equation}
hence $\displaystyle \{q_{ab}, G(\lambda)\}=0.$ Since the metric is gauge-invariant, so are all the classical observables that depend on it. Hence, for instance, the area and volume functionals are observables in this model, just as in loop quantum gravity (LQG) \cite{Rovelli:1994ge}. 

Other obvious examples of observables are $SU(2)$-valued loop variables, which are defined using traces of holonomies around curves in $\Sigma$ and insertions of configuration and momentum variables along the curves. That is, given a smooth curve $\gamma:[a,b] \to \Sigma$, we define the Wilson line from $a$ to $b$:
\begin{equation}
    U[\gamma](a,b) := \mathcal{P}\exp\left({\int_a^b dx^a A^i_a\tau_i}\right),
\end{equation}
where $\mathcal{P}$ stands for path-ordering and $\tau_i$ are the generators of $\mathfrak{su}(2)$. If $a = b$, we get the so-called Wilson loops. Since these are matrices in $SU(2)$, and hence transform under $SU(2)$ via conjugation (i.e. $g\cdot U[\gamma] = g^{-1}(a)U[\gamma]g(b)$), the cyclic invariance of the trace operation entails that
\begin{equation}
    T^0[\gamma] := \text{Tr}\,(U[\gamma](a,a)) 
\end{equation}
is an observable. Similarly, the rest of the familiar $T$ variables, defined via insertions of the momentum $\tilde{E}^a_i$ along Wilson lines, are also observables, e.g.
\begin{eqnarray*}
     T^{a_1\cdots a_n}[\gamma](x_1,\ldots, x_n) = \text{Tr}\,\left[U[\gamma](a,x_1)\tilde{E}^{a_1}(x_1) \right. \\
     \left.\times U[\gamma](x_1, x_2)\tilde{E}^{a_2}(x_2)\cdots\tilde{E}^{a_n}U[\gamma](x_n,a)\right],  
\end{eqnarray*}
where 
$\tilde{E}^a = \tilde{E}^a_i\tau^i$ and $x_1,\ldots, x_n$ are fixed points on the curve $\gamma$. 

However, unlike in   LQG, these are not the only loop observables, since we now have the additional $\mathfrak{su}(2)$-valued configuration and momentum variables, namely $\phi^i$ and $\tilde{p}_i$, that may be inserted along Wilson lines as well. There are numerous such possibilities, ranging from simple end-of-the-curve insertions such as
\begin{equation}
    \text{Tr}\,\left[\phi(x)U[\gamma](x,y)\phi(y)\right]
\end{equation}
to mixed and middle-of-the-curve insertions such as
\begin{eqnarray}
    \text{Tr}\,[\phi(x_1)U[\gamma](x_1, x_2)\tilde{E}(x_2)U[\gamma](x_2, x_3)\nonumber \\
    \left.\times\tilde{p}(x_3)U[\gamma](x_3,x_4)\phi(x_4) \right].
\end{eqnarray}
The former are essentially flux tubes analogous to those found in Yang-Mills theory, except that the nonzero Hamiltonian of the latter can cause these tubes to break to form more such tubes, whereas our model has unbroken tubes floating around in spacetime. 

\section{\label{sec4} Classical Solutions}

Since the Hamiltonian constraint, which generates orthogonal transformations of three-dimensional hypersurfaces, vanishes in both the HK model and the model here, both are essentially theories of three-geometry that arise from four-dimensional actions.  For the HK model, constraint free initial data specified on a three-dimensional hypersurface are solutions of the Gauss and spatial diffeomorphism constraints. Since the Hamiltonian constraint vanishes, these data do not evolve. Therefore they are solutions for a three-geometry. This means that constraint-free data for  Einstein gravity are a subset of the data for the HK model.

In the model at hand the Gauss constraint contains a source term. Two peculiar features of the model conspire to yield a large class of interesting three-geometries as solutions of the model. First, the three-metric (\ref{space-m}) depends solely on the configuration variables $\phi$ and $A$. Since these variables are independent, one can set their conjugate momenta $\tilde{p}^i$ and  $E^{ai}$ to zero to trivially solve the Gauss constraint, which fortunately is the only constraint to solve. Second, since the model is valid for arbitrary (semisimple) gauge groups of dimension 3, which include those with Cartan-Killing metrics of mixed signature (e.g. $SO(2,1)$), one can even construct solutions of 2+1 gravity.  Thus, by appropriately fixing the form of $\phi$ and $A$, while setting $\tilde{p}$ and $\tilde{E}$ to zero, it is possible to construct almost any three-metric, including all those resulting from projecting solutions of the 3+1 Einstein equations on three-dimensional hypersurfaces, as well as all the solutions of 2+1 gravity. To illustrate these remarks, we explicitly construct the spatial Schwarzschild metric and the Banados-Teitelboim-Zanelli (BTZ) black-hole metric \cite{Ba_ados_1992}. 

\subsection{3d Schwarzschild metric}
For $SU(2)$, $\tilde{p} = \tilde{E} = 0$, use spherical coordinates $(r, \theta, \varphi)$ on $\Sigma$, and set 
\begin{align}
    A &= \frac{r}{P(r)}\tau_1 d\theta + (\cos\theta\tau_3 - \frac{r}{P(r)}\sin\theta\tau_2)d\varphi\\
    \phi &= P(r)\tau_3,\qquad f(r) = \sqrt{1-R/r}\\ 
    P(r) &= \frac{R}{2}\left[ \ln\left|\frac{f(r)+1}{f(r)-1}\right| + \frac{2f(r)}{(f(r)+1)(f(r)-1)}\right],
\end{align}
where $\tau^i$ are the Pauli matrices. Then (\ref{space-m}) gives the  Schwarzschild three-metric
\bea
ds^2 = f^{-2}(r) dr^2  + r^2 d\Omega^2.
\eea

\subsection{BTZ black hole}
For the gauge group $SO(2,1)$ the Cartan-Killing metric can be written as $\text{diag}(-1,1,1)$. Pick cylindrical coordinates $(r, \varphi, z)$ on $\Sigma$. 
Then the BTZ metric can be obtained as follows. Let 
\bea
    A &=& [P(r)\tau_1 + Q(r)\tau_2]d\varphi + [R(r)\tau_2 + S(r)\tau_3]dz,\nn\\
    \phi &=& T(r)\tau_3, \\
    N^2(r) &=& -M+\frac{r^2}{l^2}+\frac{J^2}{4r^2}, \; N_{\varphi}(r) = -\frac{J}{2r^2}, 
\eea
where the last two functions are used to define the BTZ metric. Substituting these into (\ref{space-m}) yields the  equations
\bea
    T'(r)^2 &=& N^{-2}(r),\\
    T^2(r)(P^2(r)-Q^2(r)) &=& r^2, \\
    R^2(r)T^2(r) &=& N^2(r),\\
    R(r)Q(r)T^2(r) &=& r^2N_{\varphi}(r),
\eea
which can be solved for the functions $P,Q,R,T$ to obtain $A$ and $\phi$: the first equation gives  
\bea
    T(r) &=& \frac{l}{2}\ln\left|\sqrt{h^2(r)+1}+h(r)\right|,\nn\\
    h(r) &=& \frac{2r^2-Ml^2}{\sqrt{J^2l^2-M^2l^4}}, 
\eea
and the others three are algebraic.  

These examples show that almost any desired 3-metric may be constructed by solving for $A$ and $\phi$. 

\section{\label{sec5} Quantum Theory}
Since there are now two types of configuration variable, the connection $A$ and the $\phi$ variables, the connection representation is better termed the $A$-$\phi$ representation. Physical states are gauge-invariant functionals of $\phi$ and $A$ -- for instance, the loop variables defined above but depending solely on $A$ and $\phi$, e.g
\begin{eqnarray}
    \Psi(\phi, A) = \text{Tr}\,(\phi(x_1)U[\gamma](x_1,x_2)\phi(x_2)\nonumber \\ 
    \times U[\gamma](x_2,x_3)\phi(x_3)).
\end{eqnarray}
The inner product on these states is well-defined, with there being now an additional integration over the $\phi$:
\begin{eqnarray}
    \langle \Psi_1|\Psi_2\rangle = \int d\mu(\phi) \int d\mu(A) \Psi^*_1(\phi, A)\Psi_2(\phi, A), 
\end{eqnarray}
where $\mu(A)$ is a measure on the space of connections (modulo gauge transformations), such as the Ashtekar-Lewandowski measure, and $\mu(\phi)$ is some suitable measure on the space of the $\phi$ fields (say, a Gaussian measure). 

 
\subsection{Spin-network states}
As in LQG, quantum states in the model can be realized as spin networks, with one addition. As usual, $SU(2)$ representations label the edges of embedded graphs, corresponding to holonomies of the connection along those edges, i.e. functionals of the form $\Psi(A)$, and these representations are sewed by intertwiners associated with the vertices of the graph. However, here we can also have physical functionals that involve insertions of $\phi$ at the end of Wilson lines. In a spin network, such states correspond to the $\phi$ variables sitting at selected vertices of the underlying graph, along with the intertwiners. Thus in general, a typical spin-network state corresponding to a graph $\Gamma$ is
\begin{equation}
    |\Gamma; j_1, \ldots, j_n; I_1, \ldots, I_n; \phi_1, \ldots, \phi_k\rangle, \qquad k \leq n,
\end{equation}
with the obvious inner product. Such states are physical states of the theory, and the area and volume operators of LQG are diagonal physical observables. The quantum tetrahedra are physical states of the theory, and as there is no spatial diffeomorphism constraint, each graph representating a tetrahedron is a  distinct physical state, unlike in the HK model. 

\section{Discussion} 

Our main observation is the theory defined by (\ref{act2}), and its unusual canonical version with missing Hamiltonian and diffeomorphism constraints---it is a theory of $3$-geometries with local degrees of freedom that is exactly solvable classically and quantum mechanically.  

 A natural extension of the model arises by adding a Chern-Simons boundary term  
\begin{align}
    S &= \frac{1}{2}\int_M d^4x\,\text{Tr}\,(D\phi\wedge D\phi \wedge F) \notag\\
    &+\gamma\int_{\partial M}d^3x\,\left(A\wedge dA + \frac{2}{3}A\wedge A \wedge A\right).   
\end{align}
Varying the action now yields
\begin{widetext}
    \begin{align*}
        \delta S = -\int_M d^4x\,[D\wedge (D\phi\wedge F))]_i&\delta\phi^i + \gamma\int_{\partial M}d^3x \,\tilde{p}_i\delta\phi^i \\
        &-\int_M d^4x\, [D\wedge (D\phi\wedge D\phi) + (\phi \times (D\phi\wedge F))]^\alpha_k \delta A^k_\alpha + \int_{\partial M}d^3x\,[\gamma\tilde{\epsilon}^{abc}F_{iab} + \tilde{E}^c_k]\delta A^k_c.
    \end{align*}
\end{widetext}
Thus, if we insist that the variations in $\phi$ and $A$ on the boundary $\partial M$ can be nonzero, the conjugate momenta must be constrained on the boundary to ensure a well-defined variational principle:
\begin{equation}
    \left.\tilde{E}^c_k\right|_{\partial M} = -\gamma\tilde{\epsilon}^{abc}F_{iab},\quad \left.\tilde{p}_i\right|_{\partial M} = 0.
\end{equation}
The first of these  is a condition on conjugate momenta that also arises with a topological bulk theory \cite{Smolin:1995vq}, and in the context of black-hole entropy in LQG for an inner boundary \cite{Ashtekar_1998,PhysRevD.59.084019}, where quantization of this condition in the spin-network basis with the area operator provides an area-entropy relation. A similar calculation is possible in the present theory. 

Other possible variants of the action  include the addition of a bulk topological term $ \int d^4x\ {\rm Tr} (F\wedge F)$, more than one connection \cite{BarberoG:2019jez}, as well as boundaries with multiple components for identifying corner observables; see, e.g. \cite{Jubb:2016qzt}. 

As the model and its variants  are generally covariant theories with local degrees of freedom, they provide a useful testing ground for other quantization methods, such as those involving spinfoams \cite{Baez:1999sr,Perez:2012wv} and group field theory (GFT) \cite{Gielen:2016dss}. For the former, it is possible to write the action as a $BF$ theory with a ``simplicity" constraint term $\Lambda (B- D\phi\wedge D\phi)$; for the latter, the fact that our model has no constraints other than the Gauss law makes it a natural candidate to define directly on a group manifold, in particular for exploring the question of the manifestation of the Hamiltonian and diffeomorphism constraints in GFT models.  

 \medskip
 
\begin{acknowledgments}
This work was supported by NSERC of Canada. VH thanks the Perimeter Institute where this work was completed; Research at Perimeter Institute is supported in part by the Government of Canada through the
Department of Innovation, Science and Economic Development Canada and by the Province of Ontario through the
25 Ministry of Colleges and Universities.  
\end{acknowledgments}

\appendix
\section{Dirac analysis}
For completeness we present the Dirac analysis \cite{dirac2001lectures, Henneaux:1992ig} of the action  (\ref{act2}). This ensures consistency with the analysis of Sec. \ref{sec2}. The definitions of the momenta (\ref{eq9}) and (\ref{eq10}) have no explicit dependence on the time derivatives of $A^i_a$ and $\phi_i$ and so cannot be inverted to express the velocities as functions of the momenta and configuration variables. We thus have the following primary constraints: 
\bea
&&\tilde{E}^0_i = \frac{\p \mathcal{L}}{\p \dot{A}_0^i}\approx 0 \label{eqa1}\\
&&    \tilde{\psi}_i := \tilde{p}_i - \tilde{\epsilon}^{abc}\epsilon_{ijk}D_a\phi^jF^k_{bc} \approx 0 \label{eqa2}\\
&&   \tilde{\chi}^a_i := \tilde{E}^a_i - \tilde{\epsilon}^{abc}\epsilon_{ijk}D_b\phi^jD_c\phi^k \approx 0\label{eqa3}
\eea
Since eq. (\ref{eq11}) is in the form $\int d^3x\, \left(p\dot{q} - H\right)$, the total Hamiltonian a'la Dirac is
\begin{equation}
    H_T = \int d^3x \,(A^k_0 \tilde{G}_k + \lambda^i\tilde{E}^0_i + \mu^i\tilde{\psi}_i + \rho^i_a\tilde{\chi}^a_i),
\end{equation}
where $\lambda_i, \mu^i, \rho^i_a$ are arbitrary functions of spacetime coordinates. 

To ensure preservation of these primary constraints under evolution with $H_T$, we first note the following algebra:
\begin{widetext}
    \bea
   && \{\tilde{E}^0(x), \int d^3y\, A_0(y)\cdot\tilde{G} \} = -\tilde{G}(x)  \label{eqa5}\\
    && \{\tilde{\psi}(x), \int d^3y \,A_0(y)\cdot \tilde{G}(y)\} = -A_0(x)\times \tilde{\psi}(x) \\
    && \{\tilde{\chi}^a(x), \int d^3y\, A_0(y)\cdot\tilde{G}(y)\} = -A_0(x)\times\tilde{\chi}^a(x) \label{eqa7}\\
    && \{\tilde{\psi}(x), \int d^3y \,\mu(y)\cdot\tilde{\psi}(y)\} = \mu(x)\times D_a\tilde{B}^a(x)\\
   && \{\tilde{\psi}(x), \int d^3y\,\rho_a(y)\cdot\tilde{\chi}^a(y)\} = (\phi(x)\times\rho_a(x))\times \tilde{B}^a(x)  \\
   && \{\tilde{\chi}^a(x), \int d^3y\, \mu(y)\cdot\tilde{\psi}(y)\} = (\mu(x)\times \tilde{B}^a(x))\times\phi(x) + \mu(x) \times (\tilde{B}^a(x)\times\phi(x))\\
   && \{\tilde{\chi}^a(x),\int d^3y\, \rho_b(y)\cdot\tilde{\chi}^b(y)\} = 2\tilde{\epsilon}^{abc}[\phi(x)\times(\rho_b(x)\times D_c\phi(x))] + 2\tilde{\epsilon}^{abc}[D_c\phi(x)\times(\rho_b(x)\times\phi(x))]
\eea
where as before $\tilde{B}^{ai} = \tilde{\epsilon}^{abc}F^i_{bc}$. The first three equations are all weakly zero, while the fourth one is identically zero owing to the Bianchi identity; hence we obtain
\bea
    \{\tilde{E}_0(x), H_T\} &=& -\tilde{G}(x) \approx 0 \label{eqa11} \\
    \{\tilde{\psi}(x), H_T\} &\approx& (\phi\times\rho_a)\times \tilde{B}^a \approx 0 \label{eqa12}\\
    \{\tilde{\chi}^a(x), H_T\} &\approx&  [(\mu\times \tilde{B}^a) \times \phi] + [\mu\times(\tilde{B}^a\times\phi)] + 2\tilde{\epsilon}^{abc}[\phi\times(\rho_b\times D_c\phi)] + 2\tilde{\epsilon}^{abc}[D_c\phi\times(\rho_b\times\phi)] \approx 0
    \label{eqa13}
\eea
\end{widetext}
Eqn. (\ref{eqa11}) is a secondary constraint as its r.h.s does not involve a Lagrange multiplier; the other two equations ((\ref{eqa12}) and (\ref{eqa13})) are not secondary constraints, but merely consistency conditions on the Lagrange multipliers $\mu^i$ and $\rho^i_l$ (in line with Dirac's method \cite{dirac2001lectures, Henneaux:1992ig}). Thus the last two conditions, when set to zero, fix $\mu^i$ and $\rho^i_l$ as functions of the pahse space variables. Finally, by virtue of (\ref{eqa5}--\ref{eqa7}), $\{ G_i(x), H_T\}\approx 0$. Therefore, the constraint analysis ends, and we conclude that we have four constraints in total, the three primary constraints (\ref{eqa1}-\ref{eqa3}) and one secondary constraint (\ref{eqa11}). 

Next, we classify the four constraints into first-class or second-class. It is immediate that (\ref{eqa1}) and the Gauss law are first-class, whereas (\ref{eqa2}) and (\ref{eqa3}) are second-class. These second-class constraints  are solved strongly, which is equivalent to defining the phase space variables  $\tilde{p}_i$ and $\tilde{E}^a_i$ evident in (\ref{eqa2}) and (\ref{eqa3}). Lastly, the innocuous primary constraint (\ref{eqa1}) can be ignored, for its sole purpose is to provide a full canonical gauge generator that correctly rotates the full spacetime field $A_\alpha$ rather than just its spatial counterpart $A_a$ \cite{Castellani:1981us}. Hence we get a canonical theory defined by the momenta (\ref{eq9}) and (\ref{eq10}) subject to evolution under the Gauss law, as obtained in Section \ref{sec2}. It is worth emphasizing that the heuristic considerations in Section \ref{sec2} pertaining to the absence of any nontrivial algebraic relations among the momenta $\tilde{p}_i$ and $\tilde{E}^a_i$ and the configuration variables $\phi^i$ and $A^i_a$ are systematically reflected in the absence of any secondary second-class constraints in Dirac's procedure: the nontrivial relations (\ref{eqa12}) and (\ref{eqa13}) are not secondary constraints, but rather consistency conditions on the Lagrange multipliers. 
 
\bibliography{HK2}

\providecommand{\noopsort}[1]{}\providecommand{\singleletter}[1]{#1}%
\begin{thebibliography}{25}%
\makeatletter
\providecommand \@ifxundefined [1]{%
 \@ifx{#1\undefined}
}%
\providecommand \@ifnum [1]{%
 \ifnum #1\expandafter \@firstoftwo
 \else \expandafter \@secondoftwo
 \fi
}%
\providecommand \@ifx [1]{%
 \ifx #1\expandafter \@firstoftwo
 \else \expandafter \@secondoftwo
 \fi
}%
\providecommand \natexlab [1]{#1}%
\providecommand \enquote  [1]{``#1''}%
\providecommand \bibnamefont  [1]{#1}%
\providecommand \bibfnamefont [1]{#1}%
\providecommand \citenamefont [1]{#1}%
\providecommand \href@noop [0]{\@secondoftwo}%
\providecommand \href [0]{\begingroup \@sanitize@url \@href}%
\providecommand \@href[1]{\@@startlink{#1}\@@href}%
\providecommand \@@href[1]{\endgroup#1\@@endlink}%
\providecommand \@sanitize@url [0]{\catcode `\\12\catcode `\$12\catcode
  `\&12\catcode `\#12\catcode `\^12\catcode `\_12\catcode `\%12\relax}%
\providecommand \@@startlink[1]{}%
\providecommand \@@endlink[0]{}%
\providecommand \url  [0]{\begingroup\@sanitize@url \@url }%
\providecommand \@url [1]{\endgroup\@href {#1}{\urlprefix }}%
\providecommand \urlprefix  [0]{URL }%
\providecommand \Eprint [0]{\href }%
\providecommand \doibase [0]{http://dx.doi.org/}%
\providecommand \selectlanguage [0]{\@gobble}%
\providecommand \bibinfo  [0]{\@secondoftwo}%
\providecommand \bibfield  [0]{\@secondoftwo}%
\providecommand \translation [1]{[#1]}%
\providecommand \BibitemOpen [0]{}%
\providecommand \bibitemStop [0]{}%
\providecommand \bibitemNoStop [0]{.\EOS\space}%
\providecommand \EOS [0]{\spacefactor3000\relax}%
\providecommand \BibitemShut  [1]{\csname bibitem#1\endcsname}%
\let\auto@bib@innerbib\@empty
\bibitem [{\citenamefont {Horowitz}(1989)}]{Horowitz:1989ng}%
  \BibitemOpen
  \bibfield  {author} {\bibinfo {author} {\bibfnamefont {G.~T.}\ \bibnamefont
  {Horowitz}},\ }\href {\doibase 10.1007/BF01218410} {\bibfield  {journal}
  {\bibinfo  {journal} {Commun. Math. Phys.}\ }\textbf {\bibinfo {volume}
  {125}},\ \bibinfo {pages} {417} (\bibinfo {year} {1989})}\BibitemShut
  {NoStop}%
\bibitem [{\citenamefont {Husain}(1991)}]{Husain:1990sc}%
  \BibitemOpen
  \bibfield  {author} {\bibinfo {author} {\bibfnamefont {V.}~\bibnamefont
  {Husain}},\ }\href {\doibase 10.1103/PhysRevD.43.1803} {\bibfield  {journal}
  {\bibinfo  {journal} {Phys. Rev. D}\ }\textbf {\bibinfo {volume} {43}},\
  \bibinfo {pages} {1803} (\bibinfo {year} {1991})}\BibitemShut {NoStop}%
\bibitem [{\citenamefont {Witten}(1988)}]{Witten:1988hc}%
  \BibitemOpen
  \bibfield  {author} {\bibinfo {author} {\bibfnamefont {E.}~\bibnamefont
  {Witten}},\ }\href {\doibase 10.1016/0550-3213(88)90143-5} {\bibfield
  {journal} {\bibinfo  {journal} {Nucl. Phys. B}\ }\textbf {\bibinfo {volume}
  {311}},\ \bibinfo {pages} {46} (\bibinfo {year} {1988})}\BibitemShut
  {NoStop}%
\bibitem [{\citenamefont {Ashtekar}\ \emph {et~al.}(1989)\citenamefont
  {Ashtekar}, \citenamefont {Husain}, \citenamefont {Rovelli}, \citenamefont
  {Samuel},\ and\ \citenamefont {Smolin}}]{Ashtekar:1989qd}%
  \BibitemOpen
  \bibfield  {author} {\bibinfo {author} {\bibfnamefont {A.}~\bibnamefont
  {Ashtekar}}, \bibinfo {author} {\bibfnamefont {V.}~\bibnamefont {Husain}},
  \bibinfo {author} {\bibfnamefont {C.}~\bibnamefont {Rovelli}}, \bibinfo
  {author} {\bibfnamefont {J.}~\bibnamefont {Samuel}}, \ and\ \bibinfo {author}
  {\bibfnamefont {L.}~\bibnamefont {Smolin}},\ }\href {\doibase
  10.1088/0264-9381/6/10/001} {\bibfield  {journal} {\bibinfo  {journal}
  {Class. Quant. Grav.}\ }\textbf {\bibinfo {volume} {6}},\ \bibinfo {pages}
  {L185} (\bibinfo {year} {1989})}\BibitemShut {NoStop}%
\bibitem [{\citenamefont {Carlip}(1989)}]{Carlip:1989nz}%
  \BibitemOpen
  \bibfield  {author} {\bibinfo {author} {\bibfnamefont {S.}~\bibnamefont
  {Carlip}},\ }\href {\doibase 10.1016/0550-3213(89)90183-1} {\bibfield
  {journal} {\bibinfo  {journal} {Nucl. Phys. B}\ }\textbf {\bibinfo {volume}
  {324}},\ \bibinfo {pages} {106} (\bibinfo {year} {1989})}\BibitemShut
  {NoStop}%
\bibitem [{\citenamefont {Moncrief}(1990)}]{Moncrief:1990mk}%
  \BibitemOpen
  \bibfield  {author} {\bibinfo {author} {\bibfnamefont {V.}~\bibnamefont
  {Moncrief}},\ }\href {\doibase 10.1063/1.528950} {\bibfield  {journal}
  {\bibinfo  {journal} {J. Math. Phys.}\ }\textbf {\bibinfo {volume} {31}},\
  \bibinfo {pages} {2978} (\bibinfo {year} {1990})}\BibitemShut {NoStop}%
\bibitem [{\citenamefont {Smolin}(1992)}]{Smolin:1992wj}%
  \BibitemOpen
  \bibfield  {author} {\bibinfo {author} {\bibfnamefont {L.}~\bibnamefont
  {Smolin}},\ }\href {\doibase 10.1088/0264-9381/9/4/007} {\bibfield  {journal}
  {\bibinfo  {journal} {Class. Quant. Grav.}\ }\textbf {\bibinfo {volume}
  {9}},\ \bibinfo {pages} {883} (\bibinfo {year} {1992})},\ \Eprint
  {http://arxiv.org/abs/hep-th/9202076} {arXiv:hep-th/9202076} \BibitemShut
  {NoStop}%
\bibitem [{\citenamefont {Bakhoda}\ and\ \citenamefont
  {Thiemann}(2021)}]{Bakhoda:2020ril}%
  \BibitemOpen
  \bibfield  {author} {\bibinfo {author} {\bibfnamefont {S.}~\bibnamefont
  {Bakhoda}}\ and\ \bibinfo {author} {\bibfnamefont {T.}~\bibnamefont
  {Thiemann}},\ }\href {\doibase 10.1088/1361-6382/ac2721} {\bibfield
  {journal} {\bibinfo  {journal} {Class. Quant. Grav.}\ }\textbf {\bibinfo
  {volume} {38}},\ \bibinfo {pages} {215006} (\bibinfo {year} {2021})},\
  \Eprint {http://arxiv.org/abs/2010.16351} {arXiv:2010.16351 [gr-qc]}
  \BibitemShut {NoStop}%
\bibitem [{\citenamefont {{Husain}}\ and\ \citenamefont
  {{Kuchar}}(1990)}]{1990PhRvD..42.4070H}%
  \BibitemOpen
  \bibfield  {author} {\bibinfo {author} {\bibfnamefont {V.}~\bibnamefont
  {{Husain}}}\ and\ \bibinfo {author} {\bibfnamefont {K.~V.}\ \bibnamefont
  {{Kuchar}}},\ }\href {\doibase 10.1103/PhysRevD.42.4070} {\bibfield
  {journal} {\bibinfo  {journal} {\prd}\ }\textbf {\bibinfo {volume} {42}},\
  \bibinfo {pages} {4070} (\bibinfo {year} {1990})}\BibitemShut {NoStop}%
\bibitem [{\citenamefont {Barbero~G.}\ and\ \citenamefont
  {Villasenor}(2001)}]{BarberoG:2000dsc}%
  \BibitemOpen
  \bibfield  {author} {\bibinfo {author} {\bibfnamefont {J.~F.}\ \bibnamefont
  {Barbero~G.}}\ and\ \bibinfo {author} {\bibfnamefont {E.~J.~S.}\ \bibnamefont
  {Villasenor}},\ }\href {\doibase 10.1103/PhysRevD.63.084021} {\bibfield
  {journal} {\bibinfo  {journal} {Phys. Rev. D}\ }\textbf {\bibinfo {volume}
  {63}},\ \bibinfo {pages} {084021} (\bibinfo {year} {2001})},\ \Eprint
  {http://arxiv.org/abs/gr-qc/0012040} {arXiv:gr-qc/0012040} \BibitemShut
  {NoStop}%
\bibitem [{\citenamefont {Barbero~G.}\ and\ \citenamefont
  {Villasenor}(2010)}]{BarberoG:2010oga}%
  \BibitemOpen
  \bibfield  {author} {\bibinfo {author} {\bibfnamefont {J.~F.}\ \bibnamefont
  {Barbero~G.}}\ and\ \bibinfo {author} {\bibfnamefont {E.~J.~S.}\ \bibnamefont
  {Villasenor}},\ }\href {\doibase 10.12942/lrr-2010-6} {\bibfield  {journal}
  {\bibinfo  {journal} {Living Rev. Rel.}\ }\textbf {\bibinfo {volume} {13}},\
  \bibinfo {pages} {6} (\bibinfo {year} {2010})},\ \Eprint
  {http://arxiv.org/abs/1010.1637} {arXiv:1010.1637 [gr-qc]} \BibitemShut
  {NoStop}%
\bibitem [{\citenamefont {Husain}(1999)}]{PhysRevD.59.084019}%
  \BibitemOpen
  \bibfield  {author} {\bibinfo {author} {\bibfnamefont {V.}~\bibnamefont
  {Husain}},\ }\href {\doibase 10.1103/PhysRevD.59.084019} {\bibfield
  {journal} {\bibinfo  {journal} {Phys. Rev. D}\ }\textbf {\bibinfo {volume}
  {59}},\ \bibinfo {pages} {084019} (\bibinfo {year} {1999})}\BibitemShut
  {NoStop}%
\bibitem [{Note1()}]{Note1}%
  \BibitemOpen
  \bibinfo {note} {The condition of being semisimple is necessary, for by
  Cartan's criterion, the Lie algebra of a group is semisimple if and only if
  it possesses a non-degenerate Cartan-Killing metric, which is required to
  form a totally antisymmetric object (e.g. the Levi-Civita tensor in $\protect
  \mathfrak {su}(2)$ from the structure constants of the Lie
  algebra.)}\BibitemShut {NoStop}%
\bibitem [{\citenamefont {Dirac}(2001)}]{dirac2001lectures}%
  \BibitemOpen
  \bibfield  {author} {\bibinfo {author} {\bibfnamefont {P.}~\bibnamefont
  {Dirac}},\ }\href {https://books.google.ca/books?id=GVwzb1rZW9kC} {\emph
  {\bibinfo {title} {Lectures on Quantum Mechanics}}},\ Belfer Graduate School
  of Science, monograph series\ (\bibinfo  {publisher} {Dover Publications},\
  \bibinfo {year} {2001})\BibitemShut {NoStop}%
\bibitem [{\citenamefont {Henneaux}\ and\ \citenamefont
  {Teitelboim}(1992)}]{Henneaux:1992ig}%
  \BibitemOpen
  \bibfield  {author} {\bibinfo {author} {\bibfnamefont {M.}~\bibnamefont
  {Henneaux}}\ and\ \bibinfo {author} {\bibfnamefont {C.}~\bibnamefont
  {Teitelboim}},\ }\href@noop {} {\emph {\bibinfo {title} {{Quantization of
  gauge systems}}}}\ (\bibinfo  {publisher} {Princeton University Press},\
  \bibinfo {year} {1992})\BibitemShut {NoStop}%
\bibitem [{\citenamefont {Rovelli}\ and\ \citenamefont
  {Smolin}(1995)}]{Rovelli:1994ge}%
  \BibitemOpen
  \bibfield  {author} {\bibinfo {author} {\bibfnamefont {C.}~\bibnamefont
  {Rovelli}}\ and\ \bibinfo {author} {\bibfnamefont {L.}~\bibnamefont
  {Smolin}},\ }\href {\doibase 10.1016/0550-3213(95)00150-Q} {\bibfield
  {journal} {\bibinfo  {journal} {Nucl. Phys. B}\ }\textbf {\bibinfo {volume}
  {442}},\ \bibinfo {pages} {593} (\bibinfo {year} {1995})},\ \bibinfo {note}
  {[Erratum: Nucl.Phys.B 456, 753--754 (1995)]},\ \Eprint
  {http://arxiv.org/abs/gr-qc/9411005} {arXiv:gr-qc/9411005} \BibitemShut
  {NoStop}%
\bibitem [{\citenamefont {Banados}\ \emph {et~al.}(1992)\citenamefont
  {Banados}, \citenamefont {Teitelboim},\ and\ \citenamefont
  {Zanelli}}]{Ba_ados_1992}%
  \BibitemOpen
  \bibfield  {author} {\bibinfo {author} {\bibfnamefont {M.}~\bibnamefont
  {Banados}}, \bibinfo {author} {\bibfnamefont {C.}~\bibnamefont {Teitelboim}},
  \ and\ \bibinfo {author} {\bibfnamefont {J.}~\bibnamefont {Zanelli}},\ }\href
  {\doibase 10.1103/physrevlett.69.1849} {\bibfield  {journal} {\bibinfo
  {journal} {Physical Review Letters}\ }\textbf {\bibinfo {volume} {69}},\
  \bibinfo {pages} {1849} (\bibinfo {year} {1992})}\BibitemShut {NoStop}%
\bibitem [{\citenamefont {Smolin}(1995)}]{Smolin:1995vq}%
  \BibitemOpen
  \bibfield  {author} {\bibinfo {author} {\bibfnamefont {L.}~\bibnamefont
  {Smolin}},\ }\href {\doibase 10.1063/1.531251} {\bibfield  {journal}
  {\bibinfo  {journal} {J. Math. Phys.}\ }\textbf {\bibinfo {volume} {36}},\
  \bibinfo {pages} {6417} (\bibinfo {year} {1995})},\ \Eprint
  {http://arxiv.org/abs/gr-qc/9505028} {arXiv:gr-qc/9505028} \BibitemShut
  {NoStop}%
\bibitem [{\citenamefont {Ashtekar}\ \emph {et~al.}(1998)\citenamefont
  {Ashtekar}, \citenamefont {Baez}, \citenamefont {Corichi},\ and\
  \citenamefont {Krasnov}}]{Ashtekar_1998}%
  \BibitemOpen
  \bibfield  {author} {\bibinfo {author} {\bibfnamefont {A.}~\bibnamefont
  {Ashtekar}}, \bibinfo {author} {\bibfnamefont {J.}~\bibnamefont {Baez}},
  \bibinfo {author} {\bibfnamefont {A.}~\bibnamefont {Corichi}}, \ and\
  \bibinfo {author} {\bibfnamefont {K.}~\bibnamefont {Krasnov}},\ }\href
  {\doibase 10.1103/physrevlett.80.904} {\bibfield  {journal} {\bibinfo
  {journal} {Physical Review Letters}\ }\textbf {\bibinfo {volume} {80}},\
  \bibinfo {pages} {904} (\bibinfo {year} {1998})}\BibitemShut {NoStop}%
\bibitem [{\citenamefont {Barbero~G.}\ \emph {et~al.}(2019)\citenamefont
  {Barbero~G.}, \citenamefont {D\'\i{}az}, \citenamefont {Margalef-Bentabol},\
  and\ \citenamefont {Villase\~nor}}]{BarberoG:2019jez}%
  \BibitemOpen
  \bibfield  {author} {\bibinfo {author} {\bibfnamefont {J.~F.}\ \bibnamefont
  {Barbero~G.}}, \bibinfo {author} {\bibfnamefont {B.}~\bibnamefont
  {D\'\i{}az}}, \bibinfo {author} {\bibfnamefont {J.}~\bibnamefont
  {Margalef-Bentabol}}, \ and\ \bibinfo {author} {\bibfnamefont {E.~J.~S.}\
  \bibnamefont {Villase\~nor}},\ }\href {\doibase 10.1007/JHEP10(2019)121}
  {\bibfield  {journal} {\bibinfo  {journal} {JHEP}\ }\textbf {\bibinfo
  {volume} {10}},\ \bibinfo {pages} {121} (\bibinfo {year} {2019})},\ \Eprint
  {http://arxiv.org/abs/1906.09820} {arXiv:1906.09820 [gr-qc]} \BibitemShut
  {NoStop}%
\bibitem [{\citenamefont {Jubb}\ \emph {et~al.}(2017)\citenamefont {Jubb},
  \citenamefont {Samuel}, \citenamefont {Sorkin},\ and\ \citenamefont
  {Surya}}]{Jubb:2016qzt}%
  \BibitemOpen
  \bibfield  {author} {\bibinfo {author} {\bibfnamefont {I.}~\bibnamefont
  {Jubb}}, \bibinfo {author} {\bibfnamefont {J.}~\bibnamefont {Samuel}},
  \bibinfo {author} {\bibfnamefont {R.}~\bibnamefont {Sorkin}}, \ and\ \bibinfo
  {author} {\bibfnamefont {S.}~\bibnamefont {Surya}},\ }\href {\doibase
  10.1088/1361-6382/aa6014} {\bibfield  {journal} {\bibinfo  {journal} {Class.
  Quant. Grav.}\ }\textbf {\bibinfo {volume} {34}},\ \bibinfo {pages} {065006}
  (\bibinfo {year} {2017})},\ \Eprint {http://arxiv.org/abs/1612.00149}
  {arXiv:1612.00149 [gr-qc]} \BibitemShut {NoStop}%
\bibitem [{\citenamefont {Baez}(2000)}]{Baez:1999sr}%
  \BibitemOpen
  \bibfield  {author} {\bibinfo {author} {\bibfnamefont {J.~C.}\ \bibnamefont
  {Baez}},\ }\href {\doibase 10.1007/3-540-46552-9_2} {\bibfield  {journal}
  {\bibinfo  {journal} {Lect. Notes Phys.}\ }\textbf {\bibinfo {volume}
  {543}},\ \bibinfo {pages} {25} (\bibinfo {year} {2000})},\ \Eprint
  {http://arxiv.org/abs/gr-qc/9905087} {arXiv:gr-qc/9905087} \BibitemShut
  {NoStop}%
\bibitem [{\citenamefont {Perez}(2013)}]{Perez:2012wv}%
  \BibitemOpen
  \bibfield  {author} {\bibinfo {author} {\bibfnamefont {A.}~\bibnamefont
  {Perez}},\ }\href {\doibase 10.12942/lrr-2013-3} {\bibfield  {journal}
  {\bibinfo  {journal} {Living Rev. Rel.}\ }\textbf {\bibinfo {volume} {16}},\
  \bibinfo {pages} {3} (\bibinfo {year} {2013})},\ \Eprint
  {http://arxiv.org/abs/1205.2019} {arXiv:1205.2019 [gr-qc]} \BibitemShut
  {NoStop}%
\bibitem [{\citenamefont {Gielen}\ and\ \citenamefont
  {Sindoni}(2016)}]{Gielen:2016dss}%
  \BibitemOpen
  \bibfield  {author} {\bibinfo {author} {\bibfnamefont {S.}~\bibnamefont
  {Gielen}}\ and\ \bibinfo {author} {\bibfnamefont {L.}~\bibnamefont
  {Sindoni}},\ }\href {\doibase 10.3842/SIGMA.2016.082} {\bibfield  {journal}
  {\bibinfo  {journal} {SIGMA}\ }\textbf {\bibinfo {volume} {12}},\ \bibinfo
  {pages} {082} (\bibinfo {year} {2016})},\ \Eprint
  {http://arxiv.org/abs/1602.08104} {arXiv:1602.08104 [gr-qc]} \BibitemShut
  {NoStop}%
\bibitem [{\citenamefont {Castellani}(1982)}]{Castellani:1981us}%
  \BibitemOpen
  \bibfield  {author} {\bibinfo {author} {\bibfnamefont {L.}~\bibnamefont
  {Castellani}},\ }\href {\doibase 10.1016/0003-4916(82)90031-8} {\bibfield
  {journal} {\bibinfo  {journal} {Annals Phys.}\ }\textbf {\bibinfo {volume}
  {143}},\ \bibinfo {pages} {357} (\bibinfo {year} {1982})}\BibitemShut
  {NoStop}%
\end{thebibliography}%

\end{document}